\documentclass[aps, prb, reprint, showpacs, groupedaddress, twocolumn]{revtex4}
\usepackage{graphicx}
\usepackage{hyperref}
\usepackage{multirow}
\usepackage{amsmath}

\begin{document}

\title{Induced Work Function Changes at Mg-doped MgO/Ag(001) Interfaces: a Combined Auger Electron Diffraction and Density Functional Study}

\author{T. Jaouen}
\altaffiliation{Corresponding author.\\ thomas.jaouen@unifr.ch}
\author{P. Aebi}
\affiliation{D{\'e}partement de Physique and Fribourg Center for Nanomaterials, Universit\'e de Fribourg, CH-1700 Fribourg, Switzerland}

\author{S. Tricot}
\author{G. Delhaye}
\author{B. L{\'e}pine}
\author{D. S{\'e}billeau}
\author{G. J{\'e}z{\'e}quel}
\author{P. Schieffer}
\affiliation{D{\'e}partement Mat{\'e}riaux et Nanosciences, Institut de Physique de Rennes UMR UR1-CNRS 6251, Universit{\'e} de Rennes 1, F-35042 Rennes Cedex, France}

\begin{abstract}

The properties of MgO/Ag(001) ultrathin films with substitutional Mg atoms in the interface metal layer have been investigated by means of  Auger electron diffraction experiments, ultraviolet photoemission spectroscopy, and density functional theory (DFT) calculations. Exploiting the layer-by-layer resolution of the Mg $KL_{23}L_{23}$ Auger spectra and using multiple scattering calculations we first determine the inter-layer distances as well as the morphological parameters of the MgO/Ag(001) system with and without Mg atoms incorporated at the interface. We find that the Mg atom incorporation drives a strong distortion of the interface layers and that its impact on the metal/oxide electronic structure is an important reduction of the work function (0.5 eV) related to band-offset variations at the interface. These experimental observations are in very good agreement with our DFT calculations which reproduce the induced lattice distortion and which reveal (through a Bader analysis) that the increase of the interface Mg concentration results in an electron transfer from Mg to Ag atoms of the metallic interface layer. Although the local lattice distortion appears as a consequence of the attractive (repulsive) Coulomb interaction between O$^{2-}$ ions of the MgO interface layer and the nearest positively (negatively) charged Mg (Ag) neighbors of the metallic interface layer, its effect on the work function reduction is only limited. Finally, an analysis of the induced work function changes in terms of charge transfer, rumpling, and electrostatic compression contributions is attempted and reveals that the metal/oxide work function changes induced by interface Mg atoms incorporation are essentially driven by the increase of the electrostatic compression effect.

\end{abstract}

\date{\today}
\maketitle

\section {Introduction}
Work function is one of the most fundamental electronic properties of a metallic surface. It is the minimum energy needed to extract an electron from the solid to the infinity. Measuring the work function provides a straightforward method to monitor the state of a surface because any adsorbed species or surface defect will generally induce changes in work function. This phenomenon has been widely studied in the field of catalysis due to its important consequences concerning reaction mechanisms at surfaces because it can promote reactivity \cite{Freund2008, Surnev2013}. In particular, interfaces between ultrathin oxide films and metals are expected to play a pivotal role in controlling the charging and adsorption behaviors of metal adatoms on the oxide surface \cite{PacchioniSterrer, Benedetti2013}. Indeed, one of the major consequences of the deposition of ultrathin oxide films on metals is a shift in the metal work function resulting in a reduction of the tunneling barrier and thus an increase of the tunneling probability \cite{Jaouen2010, Pauly2010, Bieletzki2010, Konig2009, Prada2008, Goniakowski2004, Giordano2006, Butti2004}. It is now well established that these induced work function shifts can come from three distinct mechanisms: one due to the charge transfer between the oxide film and the metal substrate ($\Delta\phi^{CT}$), another due to the intrinsic dipole moment of the rumpled oxide ($\Delta\phi^{SR}$), and finally that due to the compression of the metal electronic density upon oxide deposition ($\Delta\phi^{comp}$). This last contribution, the so-called electrostatic compression effect, appears at any metal/dielectric interface and is the main mechanism governing the strong reduction of the metal work function in the case of wide band gap oxides such as MgO \cite{Goniakowski2004, Prada2008, Giordano2006}. 

The theoretical understanding of the metal/oxide interface electronic structure has thus naturally opened the way to the manipulation and the control of cluster/metal-oxide systems through interfaces engineering. By varying the kind of interface defects, it is possible to tune the work function of the metal/oxide system in a desired way \cite{Martinez2008, Martinez2009, Jerratsch2009, Wlodarczyk2012, Jaouen2012}. For examples, recent theoretical studies of doped MgO/Ag(001) interfaces, have shown that manipulating the interface by inserting oxygen or magnesium vacancies and impurities leads to an enhanced chemical reactivity of the oxide surface with respect to the catalytic dissociation of H$_2$O \cite{Jung20112012}, and have further highlighted the influence of interfacial oxygen vacancies and impurities on the MgO/Ag(001) work function \cite{Ling2013, Cho2013}. More recently, in a layer-resolved study of Mg incorporation at the MgO/Ag(001) buried interface, it has been experimentally demonstrated that Mg atoms can be incorporated at the MgO/Ag(001) interface by simple exposures of the MgO films to a Mg flux \cite{Jaouen2013}. A gradual reduction of the metal/oxide work function upon Mg exposition (up to 0.70 eV) have been further observed and attributed to band-offset variations at the interface and band bending effects in the oxide film. 

In this paper, we propose to target the underlying mechanisms responsible for the MgO/Ag(001) work function variations induced by the incorporation of Mg atoms at the metal/oxide interface. By using ultraviolet photoemission spectroscopy (UPS) and Auger electron diffraction (AED) experiment, we show that the induced work function changes are related to Fermi-level pinning modifications at the interface and we evidence a strong distortion of the Ag and MgO interface layers. The nature of interactions at the Mg-doped metal/oxide interfaces is then discussed by using density functional theory (DFT) calculations through the disentanglement of the charge transfer, rumpling, and electrostatic compression contributions.  In good agreement with our experimental observations, we found that the increase of the interface Mg concentration leads to electronic and structural changes essentially in the oxide interface layer. The strong MgO lattice distortion at the interface upon Mg incorporation is well reproduced by DFT and is found to be a structural response to the interfacial charge transfer so as to compensate almost exactly the charge transfer contribution to the work function variations, demonstrating that the metal/oxide work function changes are essentially driven by the increase of the electrostatic compression effect.

\section {Experimental details}
All experiments were performed in a multi-chamber ultrahigh vacuum (UHV) system with base pressures below 2$\times$10$^{-10}$ mBar. The (001)-oriented Ag single crystal was cleaned by several cycles of Ar$^+$ ion bombardment and annealing at 670-720 K. The 3 monolayers (ML) thick MgO films (one monolayer is defined as one half MgO lattice parameter, i.e. 1 ML = 2.10 \AA) were grown on the prepared Ag(001) surface by evaporation of Mg in O$_2$ background atmosphere (oxygen pressure = 5$\times$10$^{-7}$ mBar) at 453 K with a cube-on-cube epitaxy with respect to the Ag(001) substrate. In order to incorporate Mg atoms at the MgO/Ag(001) interface, the MgO films were exposed to an Mg flux (2.4$\times$10$^{13}$ atoms/(cm$^{2}$s))at a substrate temperature of 513 K. Under these conditions most of Mg atoms reaching the surface desorb and only a small portion of atoms is adsorbed at the film surface or incorporated. 

The measurements were carried out using x-ray and ultraviolet photoelectron spectroscopy (XPS-UPS). A two axis manipulator allowed polar and azimuthal sample rotations with an accuracy better than 0.2$^{\circ}$. Auger Electron Diffraction (AED) measurements were performed for the Mg $KL_{23}L_{23}$ Auger transition which leads to electrons with kinetic energies around 1177 eV and AED profiles were recorded during polar sample rotations (the polar angle is defined with respect to the surface normal) between -5$^{\circ}$ and 60$^{\circ}$ for the (010) and (110) inequivalent emission planes of the cubic structure of the MgO(001) film. The kinetic energy of the emitted electrons has been measured by employing a hemispherical analyzer (Omicron EA125) with a five-channel detection system. Al $K\alpha$ was used as the x-ray source and He-I resonance ($h\nu=$21.22 eV) line provided the UPS source for photoemission experiments. The total energy resolutions were respectively 0.80 and 0.15 eV for XPS and UPS. The work function of the dielectric system ($\phi_m^*$), defined as the energy of the vacuum level ($E_{Vac}$) with respect to the Fermi level of the MgO/Ag(001) system ($E_F$), is determined from the low-energy cutoff ($E_{cut}$) of the secondary photoelectron emission: $\phi_m^*= h\nu - (E_F - E_{cut})$. To this end, the samples were biased at $-8$ V to make the measurements of the very low-energy region of the spectrum reliable. 

\section {Computational method}
The multiple scattering spherical wave cluster calculations have been performed in the Rehr-Albers framework \cite{Rehr1990}, by using the MsSpec program \cite{Sebilleau2006, Sebilleau2011} for clusters containing up to 409 atoms. The multiple scattering expansion of the photoelectron wave function was carried up to the fourth order which appeared to be sufficient to achieve convergence for the considered configurations. Due to the strongly peaked shape of the scattering factor, the multiple scattering paths implying more than one scattering angle greater than 30$^{\circ}$ were neglected. Considering various experimental works \cite{Flank1996,Luches2004}, we assumed pseudomorphic ultrathin MgO films on Ag(001) with  interface Mg atoms occupying the substrate hollow sites and an interfacial distance between Ag and O atoms of 2.51 \AA. The phase shifts have been calculated within the muffin-tin approximation by using a real Hedin-Lundqvist exchange and correlation potential \cite{Hedin1969}, with muffin-tin radii of 1.02 \AA~ for O and Mg atoms and of 1.42 \AA~ for Ag atoms. The thermal vibrations were introduced by means of the isotropic mean square displacements (MSD). The MSD’s for the Mg$^{2+}$ and O$^{2-}$ ions of the different layers have been assumed identical and have been considered close to those of the MgO(001)surface ions \cite{Chen1978}. The theoretical MSD values of the Mg$^{2+}$ and O$^{2-}$ ions at the MgO(001) surface have been then averaged over all directions and both cations and anions. From this procedure we obtained a value of 0.006 \AA$^2$~ which has been introduced in our multiple scattering calculations. For the Ag bulk atoms, the MSD was 0.024 \AA$^2$, corresponding to a Debye temperature of 225 K. The MSD of the interface Ag atoms were calculated by scaling the MSD bulk value with 1.8 \cite{Yang1991, Note1}. The electron inelastic mean free path was set equal to 19 \AA~ as calculated from the Seah and Dench formula \cite{Seah1979}. Finally, a broadening of the AED peaks due to the formation of mosaic observed during the growth of the MgO films on Ag(001) \cite{Wollschlager1998},was taken into account by averaging the calculations over directions filling a cone of 2.5$^{\circ}$ half angle.

\begin{figure*}[t]
\includegraphics[width=0.96\textwidth]{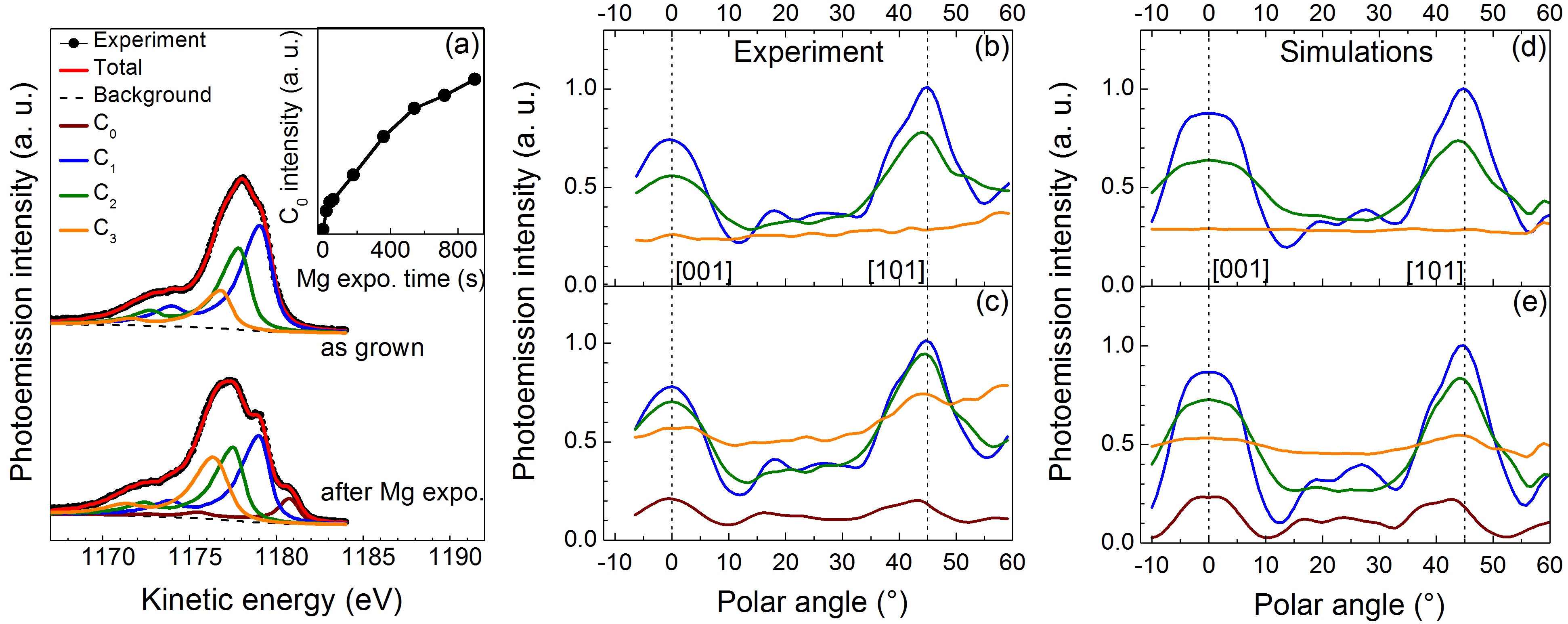}
\caption{\label{fig1}(color online). (a) Photoemission spectra of the Mg $KL_{23}L_{23}$ Auger transition of the 3 ML MgO film obtained before and after exposition to an Mg atomic flux. Best fit and layer-by-layer decomposition are also shown. The inset shows the evolution of the intensity of the $C_0$ component as a function of the Mg exposure time. (b)-(c) Experimental AED polar scans of the $C_1$, $C_2$, and $C_3$ Auger components in the (010) emission plane, for the reference sample (b) and of the $C_0$, $C_1$, $C_2$, and $C_3$ Auger components in the (010) emission plane, after Mg exposure (c).(d)-(e) Calculated Mg $KL_{23}L_{23}$ AED profiles in the (010) emission plane for a 3 ML MgO system (d) and with Mg atoms occupying substitutional sites of the Ag substrate (e).}
\end{figure*}

The DFT calculations have been carried out in the generalized gradient approximation (GGA) using the Perdew-Burke-Ernzerhof (PBE) exchange-correlation functional.\cite{Perdew1996} All calculations were performed within the Projector-Augmented Wave (PAW) formalism \cite{Blochl1994}, implemented in a real-space grid in the GPAW code \cite{Enkovaara2010, Mortensen2005}, with a grid spacing of 0.18 \AA. The MgO/Ag (001) system was modeled with one, two or three-layers of MgO deposited on three Ag layers with lattice parameter $a_0=$ 4.16 \AA~ and Ag interface atoms below the oxygen anions. The DFT-optimized Ag lattice constant ($a_0=$4.16 \AA) being 2\% smaller than the MgO one ($a_0=$ 4.25 \AA), the MgO layers are slightly contracted with respect to their bulk distance when supported on the Ag metal. During geometry optimization of the MgO/metal interface, all atoms in the MgO film and in the two-interface nearest Ag layers were relaxed until the atomic forces are less than 0.02 $eV.\AA^{-1}$ per atom while the remaining metal layer was frozen at bulk positions. In order to study the impact of the Mg incorporation on the physical properties of the MgO/metal system, Ag atoms were replaced by Mg atoms in the substrate interface layer (structure labeled MgO/Ag-Mg in the remainder of the paper). Surface unit cells with $(\sqrt{2}\times\sqrt{2})a_0$ dimension were used for calculations and Brillouin zone integration was performed using 4$\times$4$\times$1 Monkhorst-Pack meshes \cite{MonkhorstPack1976} To improve the convergence we used a Methfessel-Paxton occupation function with $k_BT=$0.2 $eV$. The vacuum region between adjacent slabs was set to $\sim$20 \AA~ and dipole correction was applied in order to calculate the work functions. The convergence study of the work function has been done for a clean Ag(001) surface as a function of the Ag slab thickness and has shown that the work function value was converged within $\sim$3 \% between 3 and 8 Ag layers. The charges of the different atoms have been obtained through a Bader analysis.\cite{Tang2009}
  
\section {Auger electron diffraction and photoemission analysis}
\subsection {AED}

Figure \ref{fig1}(a) shows normal-emission Mg $KL_{23}L_{23}$ Auger spectra (vertically shifted for clarity) of the 3 ML MgO sample before and after an exposition to a Mg atomic flux (2.4$\times$10$^{13}$ atoms/(cm$^{2}$s)) during 12 minutes at a substrate temperature of 513 K. The 3 ML reference spectrum is fitted by three Auger components $C_1$, $C_2$, and $C_3$,  with maxima situated at 1179.1 eV, 1177.8 eV, and 1176.8 eV and reveals a layer-resolved shift for the Mg $KL_{23}L_{23}$ Auger transition. Indeed, as demonstrated in Ref. \onlinecite{Jaouen2013}, the $C_1$, $C_2$, and $C_3$ components correspond respectively to Auger electron emission from the first, second and third MgO plane above the metal-oxide interface for a 3ML thick MgO film. After Mg exposure, a fourth component (labeled $C_0$) grows at higher kinetic energy in the Mg $KL_{23}L_{23}$ Auger spectrum (1180.8 eV). This component whose intensity is comparable to the other contributions and the kinetic energy very close to that obtained for a fraction of a Mg monolayer deposited on Ag(001), is a spectroscopic fingerprint of electron emission from Mg atoms located in a metallic environment. More precisely, the component $C_0$  is related to an electron emission from Mg atoms incorporated in the Ag substrate during the Mg exposure \cite{Jaouen2013}. As it can be seen in the inset of Fig. \ref{fig1}, its normal-emission intensity gradually increases as a function of the Mg exposition time. This suggests that the incorporation mechanism takes place throughout the Mg exposure.

Figures \ref{fig1}(b) and \ref{fig1}(c) respectively show the experimental AED polar scans in the (010) emission plane associated with each fitting component of the  Mg $KL_{23}L_{23}$ Auger spectra for the reference sample and the same sample after the Mg exposure. The curves were obtained from the intensities extracted from the fitting procedure and divided by the same factor which has been chosen such that the $C_1$ AED curve is normalized with its intensity equal to 1. As the $C_1$, $C_2$, and $C_3$ components of the as-grown Mg $KL_{23}L_{23}$ Auger spectrum correspond respectively to Auger emissions from the first, second and third MgO layer above the metal-oxide interface, the intensity distribution of the $C_3$ is isotropic whereas the AED profiles associated with the $C_1$ and $C_2$ contributions show pronounced peaks at normal emission and at 45$^{\circ}$. In the forward focusing approximation \cite{Egelhoff1990}, these maxima correspond respectively to scattering events along the [001] and [101] atomic directions of the rocksalt (NaCl) structure of the MgO lattice.  

We have then exploited the layer-by-layer resolution of the Mg $KL_{23}L_{23}$ Auger transition to extract the structural parameters of the MgO film by means of multiple scattering calculations with the MsSpec program \cite{Sebilleau2006, Sebilleau2011}. For the calculations we have considered a pseudomorphic MgO film with an in-plane lattice parameter of 4.09 \AA~ that fully covers the Ag(001) surface, neglecting the MgO rumpling distortion (the amplitude of the rumpling is generally no more than few percent of interplanar distance in MgO) and fixing the distance $d_{Ag-0}$ (= 2.51 \AA) between Ag and O atoms at the MgO/Ag(001) interface. By comparing the experimental and calculated intensities (averaged over all polar angles in the (010) emission plane) (see Fig.\ref{fig1}(b)-(d)), it is found that the real MgO coverage is 2.6 ML with bilayer and trilayer occupancies of $\sim$37\% and $\sim$63\%, respectively. 

A quantitative comparison of the calculated and measured angular positions of the forward focusing peaks along the [101] direction was carried out to determine the inter-planar distance $d_{ij}$ between two successive $i$ and $j$ MgO layers (the index $i$=1 refers to the MgO interface layer). The various parameters deduced from our analysis are summarized in Table \ref{tab1}. The inter-planar distances are very close to those found by Luches \textit{et al.} (2.14-2.15 \AA) by using Extended X-ray absorption fine structure (EXAFS) on a 3 ML MgO/Ag(001) sample \cite{Luches2004}. The calculated AED curves taking into account the structural and morphological parameters given on the left hand side of Table \ref{tab1} are shown in Fig.\ref{fig1}(d). Note the remarkable agreement between the positions and the shapes of the experimental and calculated AED peaks for the different Auger components. 

\begin{table}[t]
\caption{\label{tab1}Structural and morphological parameters of the MgO(3ML)/Ag(001) samples before (3 ML MgO) and after Mg exposition (After Mg expo.): distance between the incorporated Mg atoms and the O$^{2-}$ ions of the MgO interface layer $d_{Mg-O}$, inter-planar distances $d_{ij}$ between two successive $i$ and $j$ MgO planes, proportion of bilayer, trilayer, and quadrilayer in the MgO films (Occupancy), and coverage of the Ag(001) surface.}
\begin{ruledtabular}
\begin{tabular}{lcccc}
\multirow{2}{*}{} & \multicolumn{2}{c}{3 ML MgO}& \multicolumn{2}{c}{After Mg expo.}\\ \cline{2-3} \cline{4-5}
& Bilayer & Trilayer & Trilayer & Quadrilayer\\ \hline
$d_{Mg-O}$(\AA) & & & 2.44$\pm{0.08}$\footnote{The same distances were considered for the MgO trilayer and quadrilayer} & 2.44$\pm{0.08}$\footnotemark[1]\\
$d_{12}$(\AA) & 2.12$\pm{0.04}$& 2.13$\pm{0.07}$& 2.22$\pm{0.06}$\footnotemark[1] & 2.22$\pm{0.06}$\footnotemark[1]\\
$d_{23}$(\AA) & & 2.14$\pm{0.04}$ & 2.14\footnote{Fixed distances} & 2.14\footnotemark[2]\\
$d_{34}$(\AA) & & & & 2.14\footnotemark[2]\\
Occupancy (\%) & 37$\pm{10}$ & 63$\pm{10}$ & 83$\pm{10}$ & 17$\pm{10}$\\
Coverage (\%) &\multicolumn{2}{c}{100}& \multicolumn{2}{c}{89}\\
\end{tabular}
\end{ruledtabular}
\end{table}

Let us now discuss the differences between the AED curves of the $C_1$, $C_2$, and $C_3$ components obtained before and after the Mg exposure. The $C_1$ and $C_2$ AED curves stay almost unchanged, while the post-growth Mg exposure leads to an enhancement of the $C_3$ contribution with the appearance of weak intensity modulations at 0$^{\circ}$ and 45$^{\circ}$ (Fig. \ref{fig1}(c)) indicating that the MgO film is slightly reorganized with a small fraction of the oxide thicker than 3 ML. Assuming that the maximum thickness of the film does not exceeds 4 atomic layers and comparing the experimental data with the calculated intensities, we find that $\sim$83\% and $\sim$17\% of the MgO film are respectively in trilayer and quadrilayer configuration and that the MgO film covers $\sim$89\% of the Ag(001) surface. The $C_0$  metallic component curve in Fig. \ref{fig1}(c) shows a well-structured pattern with forward scattering peaks along the [001] and [101] directions similar to those observed for $C_1$ and $C_2$ components. Also, we can note that the forward-focusing peak along the [001] direction is sharper than those of the $C_1$ and $C_2$ components. Considering the defocusing effects related to the multiple scattering events, such a narrowing is expected for electron emission from Mg atoms located beneath the oxide film \cite{Xu1989, Aebischer1990}. This confirms that the $C_0$  component is associated with Auger electron emission from Mg atoms in the Ag substrate. 

However, the forward scattering peaks along the low index directions at 0 and 45$^{\circ}$ are still well pronounced indicating that the multiple scattering defocusing effects are not sufficient to destroy them.  In other words the Mg atoms in metallic environment are located in the vicinity of the MgO/Ag interface. In fact, the multiple scattering calculations show that the $C_0$ component is related to electron emission from Mg atoms preferentially incorporated in the substitutional sites of the Ag plane just beneath the MgO lattice. For an Mg emitter located in the second layer of the Ag substrate, the intensity of the forward focusing peak around 45$^{\circ}$ is much smaller than that of the peak at 0$^{\circ}$. By varying the multiple scattering order we have found that the defocusing effect along the [101] direction is more efficient than along the [001] direction because there is an additional Ag scatterer along the [101] direction. A fortiori, the multiple scattering defocusing effects will destroy the forward-focusing structure for the Auger emission from Mg atoms deeper into the Ag substrate and lead to the emergence of dominant structures in the intermediate directions which do not correspond to low index directions of the system.

\begin{figure}[b]
\includegraphics[width=0.48\textwidth]{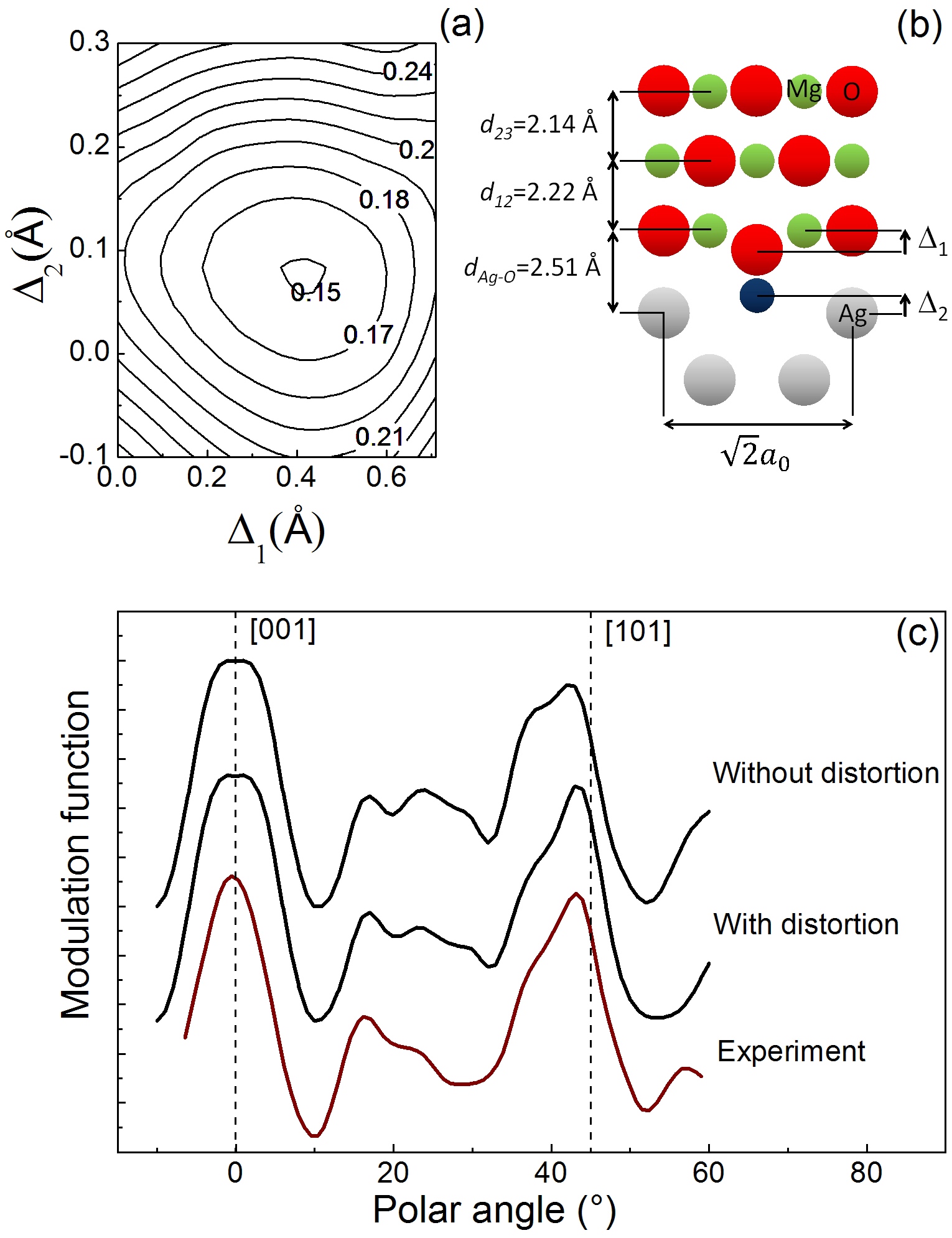}
\caption{\label{fig2}(color online). (a) $R_p$-factor contour map using $\Delta_1$ and $\Delta_2$ as parameters. These parameters are defined in (b) which shows a sketch of our structural model. The red, green and grey atoms correspond respectively to oxygen, magnesium and silver. The incorporated magnesium emitter atom is shown in blue.(c) Comparison between experimental (bottom curve) and calculated modulation functions (top curves) associated with the $C_0$ component in the (010) emission plane for Mg emitter atoms located in the first substrate layer with or without lattice distortion. }
\end{figure}

The structural parameters of the MgO film after the Mg exposure have also been determined using multiple scattering calculations. The inter-layer distances have been deduced in the same manner as above, i.e. by a quantitative comparison of the calculated and measured angular positions of the forward focusing peaks along the [101] direction. For these calculations, we have fixed the inter-planar distances $d_{23}$ and $d_{34}$ to 2.14 \AA, we have assumed a pseudomorphic MgO film on Ag(001) and we have neglected the rumpling effects in the oxide film. For the Mg atoms incorporated in the Ag substrate we have considered that they occupy the substitutional sites of the Ag plane in contact with the MgO film with an atomic concentration of 25\% and an (2$\times$2) order. No rumpling was introduced in the interface alloy plane and we have assumed the same $d_{12}$ and $d_{Mg-O}$ distances for the trilayer and quadrilayer configurations ($d_{Mg-O}$ is the distance between the incorporated Mg atoms and the O$^{2-}$ ions of the MgO interface layer). The parameters deduced from our analysis are given on the right hand side of Table \ref{tab1}. We find, with our structural model, that the inter-planar distances $d_{12}$ are only slightly modified by the Mg incorporation.
As it can be seen in Fig. \ref{fig1}(c) and \ref{fig1}(e) we obtain a satisfactory agreement between the positions and the shapes of the experimental and calculated AED peaks for the different Auger components.

A closer investigation of the $C_0$ modulations demonstrates that we can improve the agreement between experiments and multiple scattering calculations by taking into account the lattice distortion of the interface layers. Such a distortion upon Mg incorporation has been predicted by our DFT-based geometry optimization calculation (see section V) that shows that the MgO interface layer undergoes a significant distortion when an Mg atom is incorporated into the Ag interface plane. In particular, we found that the O$^{2-}$ ions located directly above the Mg atom are displaced towards the Ag substrate and that a slight atomic corrugation appears in the Ag-Mg interface alloy. The physical origins of the distortion effect will be discussed in detail in section V.

Based on the previous simulations we have thus carried out a study of the MgO lattice distortion upon Mg incorporation combining multiple scattering simulations and reliability factor ($R_p$-factor) analysis \cite{Rfactor} (Fig. \ref{fig2}). The calculations were performed with a substrate containing 50\% of Mg atoms in the first substrate layer in a checkerboard arrangement, and the $R_p$-factor was calculated from AED data of the (010) and (110) emission planes. Whereas the $d_{Ag-O}$ distance was fixed to 2.51 \AA, the vertical position of the incorporated Mg atoms and the nearest neighbor O$^{2-}$ ions were modified in the optimization procedure. We find that the $R_p$-factor can be diminished by $\sim$30\% when the lattice distortion is taken into account in the calculations. 

The $R_p$ factor contour map is shown in Fig. \ref{fig2}(a) ($\Delta_1$ and $\Delta_2$ parameters are defined in the sketch of our structural model Fig. \ref{fig2}(b)). Integrating the morphological parameters of Table \ref{tab1}, the best agreement between experimental and calculated AED profiles is obtained for $d_{Mg-O}$= 2.0 \AA~ and a $R_p$-factor of 0.15. The Mg atoms of the interfacial alloy are displaced toward the oxide layer by 0.1 \AA~ and the nearest neighbors O$^{2-}$ ions are displaced downward by 0.4 \AA~ relatively to the Mg$^{2+}$ ions position. Fig. \ref{fig2}(c) shows the $C_0$ experimental modulation function $\chi$ in the (010) emission plane compared with those calculated for Mg emitter atoms located in the first substrate layer with or without lattice distortion and taking into account the morphological parameters given in Table \ref{tab1} (the modulation functions $\chi$ correspond to normalized experimental or calculated AED curves forced to have an amplitude between $-$0.5 and 0.5). Clearly, the introduction of the lattice distortion in the multiple scattering calculations improves the agreement between experimental and simulated curves.

In the present work we demonstrate the possibility to incorporate Mg atoms at the MgO/Ag interface by a simple MgO film exposure to a Mg flux. Through a quantitative analysis of the AED data we find that the atomic concentration of Mg atoms in the interfacial Mg-Ag alloy is about 30\% after an exposition of 12 minutes at a substrate temperature of 513 K. A longer exposition time at the same temperature leads to a progressive increase of the Ag 3d$_{5/2}$ core level binding energy as a function of the time indicating the formation of a bulk-like Ag-Mg alloy. Besides, the $C_0$  modulation curve shows similitudes with the one obtained for the Ag 3d$_{5/2}$ core-level. This indicates that the Mg atoms preferentially occupy the substitutional sites of the Ag lattice. Taking as reference the x-ray photoelectron spectroscopy measurements for ordered Ag$_3$-Mg and Ag-Mg alloys from Liu \textit{et al.} \cite{Liu1994}, and an Ag 3d$_{5/2}$ core level binding energy shift of 0.1 eV measured after an exposition of 38 minutes, we obtain a mean Mg atomic concentration of $\sim$10-15\% for the bulk-like Ag-Mg alloy.
 
Concerning the incorporation mechanism, it is unlikely that Mg atoms go through non-defective MgO layers at 513 K. Indeed, the diffusion coefficient associated to Mg self-diffusion in bulk-MgO is expected to be very low in this temperature range \cite{Sakaguchi1992}. Moreover, as shown by first-principle calculations \cite{Geneste2005}, Mg atoms are weakly bound to oxygen on flat MgO(001) surface and can easily desorb at the temperature we used. In contrast, they bind rather strongly to the low-coordinated sites of MgO surfaces \cite{Kantorovich1999}. Hence, we believe that Mg atoms arriving at the MgO surface diffuse until reaching defect sites. The Mg atom diffusion may then take place through low-energy pathways such as dislocations before being incorporated within the Ag interface plane. We can also note that after the Mg exposure at 513 K the MgO film covers only 89\% of the Ag substrate. Thus, the Mg atoms reaching the bare Ag substrate may also be incorporated within the Ag surface and then could migrate to the MgO/Ag interface. 

During the annealing stage under an Mg flux we have observed both a diminution of $\sim$11 \% of the substrate coverage and the occurrence of MgO quadrilayer’s. Such a morphological evolution could drive a structural evolution of the film and considering the works of S. Valeri \textit{et al.} \cite{Valeri2002}, in which it has been found the onset of misfit dislocations between 3 and 6 ML of MgO, we can expect to observe the strain relaxation through the insertion of misfit dislocations within the MgO quadrilayer. Note that the formation of dislocation-free islands also could contribute to strain relaxation. However, as the difference between the lattice parameters of MgO and Ag lattices is small, the in-plane lattice parameter of the MgO films must be close to the one of the silver substrate for the considered MgO thicknesses ($\leq$ 4 ML). At this stage, additional studies based on scanning tunneling microscopy experiments could be highly desirable to get more insights on the incorporation processes and on the structural properties of the MgO films.
	   
\subsection {UPS}

\begin{figure}[b]
\includegraphics[width=0.48\textwidth]{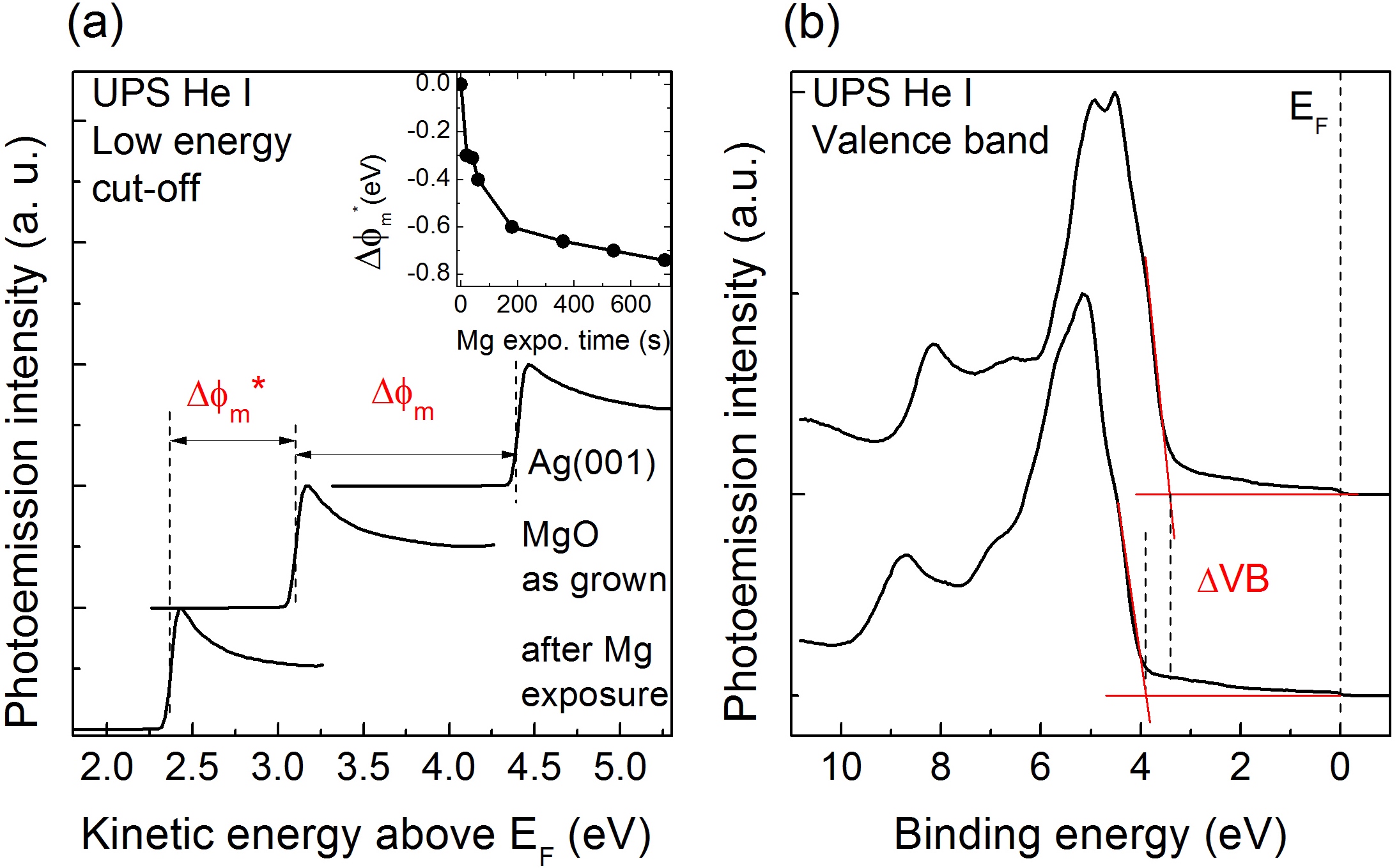}
\caption{\label{fig3}(color online).(a) Low-energy cutoff of the secondary photoelectrons emission for the Ag(001) substrate and for the MgO(3ML)/Ag(001) sample before and after exposure to a Mg flux. $\Delta\phi_m$ and $\Delta\phi_m^*$ correspond to the work function variations of the metal substrate and of the metal/oxide system, respectively. The inset shows the evolution of $\Delta\phi_m^*$ as a function of the Mg exposure time. (b) He-I UPS spectra showing the valence band region of the MgO reference sample (top) and after Mg exposure (bottom). The energy reference is taken at the Fermi level $E_F$. The method used for the VBM position determination is also sketched.}
\end{figure}

It is expected that the presence of Mg interfacial atoms should lead to strong interface electronic structure changes. Fig. \ref{fig3}(a) shows the low-energy cutoff of the secondary photoelectrons emission for the Ag(001) substrate and for the MgO(3ML)/Ag(001) sample before and after exposure to a Mg flux. The energy reference is taken at $E_F$ and the low-energy cutoff positions were taken at the maximum slope of their rising edges. The MgO deposition leads to a metal work function shift, $\Delta\phi_m$, of about -1.30 $\pm$ 0.05 eV which results in a metal/oxide work function value $\phi_m^*$ of 3.10 $\pm$ 0.05 eV for our reference sample. This strong decrease is mainly driven by the so-called electrostatic compression effect \cite{Jaouen2010}, and is in very good agreement with previous results of the literature \cite{Jaouen2010, Bieletzki2010, Konig2009}. After Mg flux exposure, we observe a stronger decrease of the metal/oxide work function reaching about $\Delta\phi_m^*=-0.70 \pm 0.05$ eV. Fig. \ref{fig3}(b) shows the valence-band (VB) region of the He-I UPS spectra corresponding to the reference MgO/Ag(001) sample (top) and after Mg exposure (bottom). The experimental value of the position of MgO valence band maximum (VBM) with respect to $E_F$ is obtained by a linear extrapolation of the leading edge of the VB spectrum and is determined to be 3.90 $\pm$ 0.10 eV and 3.40 $\pm$ 0.10 eV before and after Mg exposure, respectively. 

From our UPS measurements, we thus see that the metal/oxide work function variation differs from the one of the VBM. As shown in our previous paper \cite{Jaouen2013}, this difference comes from an initial band bending of about 0.30 eV resulting from a positive charge accumulation on the MgO surface. Indeed, during the treatment, the Mg atoms which have been directly incorporated at defects sites such as kinks or step edges lead to the creation of color centers. As recently discussed by Ling \textit{et al.} in a DFT study of the MgO/Ag(001) interface \cite{Ling2013}, the $F^0$ and $F^+$ color centers associated with oxygen vacancies on low-coordinated sites at the surface are in fact highly stable. The $F^+$ center creation at the surface thus induces an electron transfer from the oxide layer to the metal substrate and results in a downward band bending in the MgO film. Then, further Mg exposition leads to Mg diffusion up to the MgO/Ag(001) interface thereby changing the Fermi level pinning position in the MgO band gap over 0.50 $\pm$ 0.10 eV depending on the incorporated Mg concentration. 

In addition to the work function variation of the metal substrate caused by the interface alloy formation ($\Delta\phi_m^{alloy}$), the work function variation of the metal/oxide system ($\Delta\phi_m^*$), defined as the difference between the work functions of the MgO/Ag-Mg ($\phi^{MgO/Ag-Mg}$) and MgO/Ag ($\phi^{MgO/Ag}$) systems, can come from three distinct mechanisms that are the mechanism of charge transfer ($\Delta\phi^{CT}$), the mechanism of structural relaxation within the oxide film ($\Delta\phi^{SR}$), and the electrostatic compression effect ($\Delta\phi^{comp}$). In the present paper, we propose to decompose $\Delta\phi_m^*$ in a similar manner as Prada \textit{et al.}:\cite{Prada2008}

\begin{align}
\Delta\phi_m^{*}&=\phi^{MgO/Ag-Mg}-\phi^{MgO/Ag} \notag \\
 &=\Delta\phi^{comp}+\Delta\phi^{CT}+\Delta\phi^{SR}+ \Delta\phi_m^{alloy}
\end{align}

In order to determine the physical origin of the work function change induced by the Mg incorporation at the MgO/Ag(001) interface we will attempt, in the remainder of this paper, to quantify the relative importance of these effects with the help of DFT calculations.

\section {Discussion and density functional theory results}

\begin{figure}[t]
\includegraphics[width=0.48\textwidth]{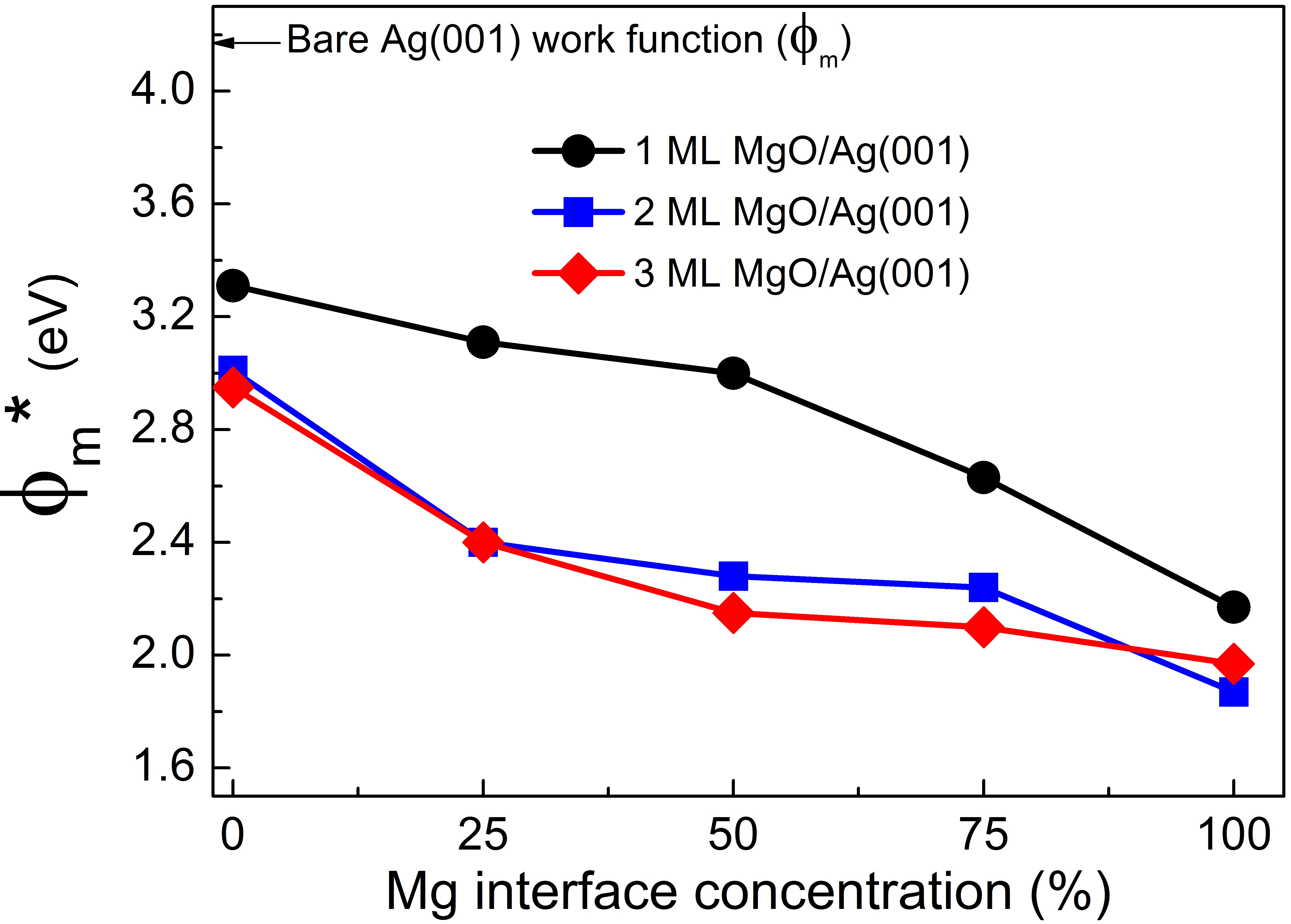}
\caption{\label{fig4}(color online). Evolutions of the calculated MgO/Ag(001) work function $\phi_m^*$ as a function of the Mg atom concentration in the Ag interface plane for oxide thickness's of 1 ML, 2 ML and 3 ML.}
\end{figure}

Figure \ref{fig4} shows the evolutions of the calculated MgO/Ag(001) work function $\phi_m^*$ as a function of the Mg atom concentration in the Ag interface plane for oxide thicknesses of 1, 2, and 3 ML. First, by focusing on the case of the clean MgO/Ag(001) interface (see 0\% Mg concentration in Fig. \ref{fig4}), we can see that the DFT-calculated Ag work function ($\phi_m=$4.17 $eV$) strongly decreases (to 3.31 $eV$, $\Delta\phi_m=$-0.86 $eV$) when adding 1 ML of MgO. This reduction is lower after deposition of a second oxide layer ($\Delta\phi_m=$-1.16 $eV$), and no more changes are observed beyond this thickness ($\Delta\phi_m=$-1.22 $eV$). Most of the variation of the work function is due to the addition of a single oxide layer (-0.86 $eV$) and is related to the highly ionic character of MgO whose interaction with the metal substrate is dominated, already for 1 ML, by electrostatic effects.\cite{Prada2008} Note that these work function reductions are in excellent agreement with previous calculations,\cite{Prada2008, Goniakowski2004, Giordano2006, Butti2004, Giordano2006bis, Sementa2012, Ling2013, Cho2013}  and very well reproduce experimental results of the literature.\cite{Jaouen2010, Bieletzki2010, Konig2009} 

Next, whatever the oxide thickness, the incorporation of Mg atoms at the interface leads to a decrease in the MgO/Ag(001) work function. The variations are almost identical for 2 and 3 ML MgO and very different from those observed for 1 ML. This indicates that most of the changes in electronic and structural properties of the MgO/Ag (001) concern the two first MgO layers starting from the interface. Note already that for 3ML MgO the change in work function calculated for a 25\% interface Mg atom concentration (-0.55 $eV$) is very close to the one determined experimentally by UPS. 

\subsection {Ag-Mg alloy work function}

Let us first consider the impact of Mg atoms alloyed with the Ag(001) on the work function $\phi_m$ of the substrate itself. In Table \ref{tab2} are given the Ag(001) work function variations $\Delta\phi_m^{alloy}$ induced by Mg atoms located in the substitutional site of the first Ag layer for an Ag-Mg substitutional surface alloy on Ag(001) after structure optimization (Ag-Mg relaxed) and for the same alloy with an atomic structure extracted from the geometry optimization of a MgO(3ML)/Ag-Mg system, but after removal of the oxide film (Ag-Mg fixed). 

\begin{table}[t]
\caption{\label{tab2}Dependences of the Ag(001) work function variation $\Delta\phi_m^{alloy}$ on the alloyed Mg atom concentration for the relaxed structure(Ag-Mg relaxed) and the fixed structure extracted from the optimization of the 3 ML MgO/Ag-Mg system (Ag-Mg fixed).}
\begin{ruledtabular}
\begin{tabular}{ccc}
\multirow{2}{*}{Mg int.(\%)} & \multicolumn{2}{c}{$\Delta\phi_m^{alloy}$ (eV)}\\ \cline{2-3}
& Ag-Mg relaxed & Ag-Mg fixed\\ \hline
0 & 0 & 0\\
25 & $-$0.05 & $-$0.11\\
50 & $-$0.09 & $-$0.17\\
75 & $-$0.25 & $-$0.29\\
100 & $-$0.31 & $-$0.35\\
\end{tabular}
\end{ruledtabular}
\end{table}

As we can see the changes associated with the two geometries are low and almost the same, i.e. they show an identical quasi-linear decrease of low amplitude as a function of the Mg concentration. Thus, the similarity of the changes associated with these two structures indicates that the structural changes related to the presence of MgO play a small role in the variation of the work function of the Ag(001) substrate induced by the alloy formation. From the present analysis we thus conclude that the work function reduction induced by the Mg incorporation cannot be simply explained by the modification of the substrate work function $\phi_m$. 

\subsection {Charge transfer}

\begin{figure}[b]
\includegraphics[width=0.48\textwidth]{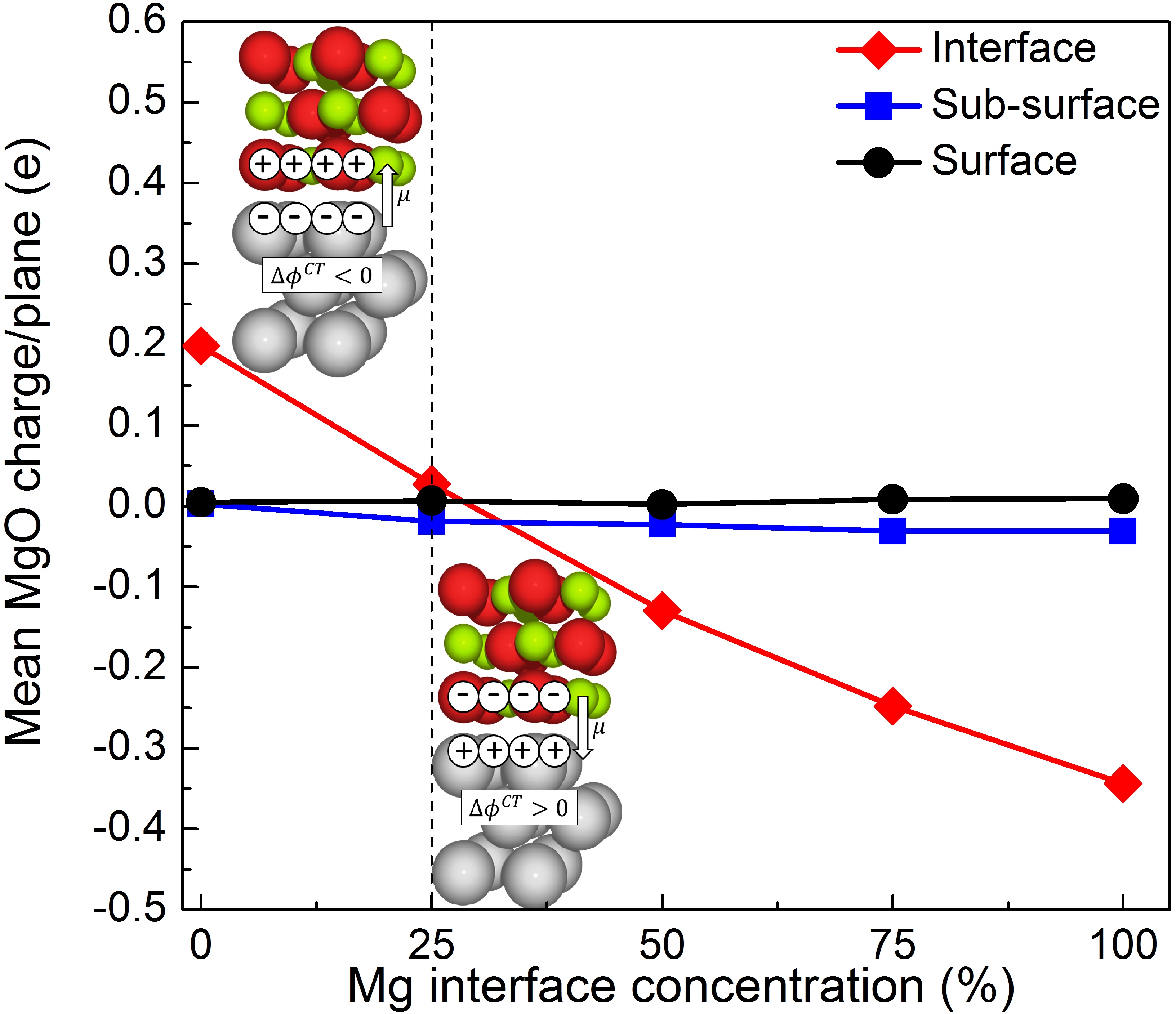}
\caption{\label{fig5}(color online). Evolutions of the mean Bader charges (where e is the elemental charge) of the MgO layers of the MgO(3ML)/Ag(001) system as a function of the Mg atom concentration in the Ag interface plane.}
\end{figure}

Figure \ref{fig5} shows the evolution of the mean charges per MgO plane and unit cell of the interface, sub-surface, and surface layers of the MgO(3ML)/Ag(001) system as a function of the Mg atom concentration in the Ag interface plane. The mean value has been obtained, after geometry optimization, through a Bader analysis by the determination of the total charge per MgO unit cell used in our DFT calculation that includes four oxygen and magnesium ions per layer. We note that depending on the Mg concentration only the charge of the MgO interface layer shows important variations.

The charge variations observed for the surface and sub-surface layers are negligible and these layers are close to the electrical neutrality. The charge transfer contribution to the work function variations is thus mainly absorbed by the MgO interface layer. The interface mean charge decreases almost linearly with the Mg concentration and presents a sign change for a Mg concentration of 25\%. In contrast to the clean MgO/Ag(001) interface for which the relatively low charge transfer (electron transfer) occurs from the oxide layer to the metal substrate, the augmentation of the interface Mg concentration leads to an increasing electron transfer from the metal to the MgO which results in a negative net charge for the oxide interface layer above Mg concentration of 25\%. The presence of these negative net charges above the metal surface leads to an increase of the surface dipole and, consequently, to work function augmentation, which does not explain our observations (see below).   

In an \textit{ab initio} theoretical study, Goniakowski and Noguera found that there is no charge transfer between a metal substrate and MgO layer for a Pauling electronegativity of the metal substrate around 1.7 \cite{Goniakowski2004}. Given that the Pauling electronegativity of Mg and Ag are 1.31 and 1.93 respectively, a sign change of the charge transfer is expected with the increasing of the Mg concentration. Assuming that the geometric mean of the Pauling electronegativities for the planar Ag-Mg alloy is sufficient to describe the substrate electronegativity, we found that the sign change must appear for a Mg concentration around 30\%, a value very close to that obtained in Fig. \ref{fig5}. 

In order to obtain the charge transfer contribution to the total work function change, we have used the procedure proposed by Prada \textit{et al.} for MgO/Ag(001) and MgO/Au(001) interfaces.\cite{Prada2008} In their study, the charge transfer contributions to the work function variations of the Ag(001) and Au(001) substrates have been determined to be -0.23 $eV$ and -0.61 $eV$, respectively for charge transfer per unit of surface ($CT/S$) of 0.65$\times10^{-2}e/\AA^2$ and 1.46$\times10^{-2}e/\AA^2$. By considering a linear relationship between $\Delta\phi^{CT}$ and $CT/S$ and by using the values of Prada \textit{et al.} to determine it, we can obtain $\Delta\phi^{CT}$ for any Mg concentration at the MgO/Ag(001) interface (Table \ref{tab3}).

\begin{table}[t]
\caption{\label{tab3}Evolutions of the charge transfer per unit cell $CT$, per unit of surface $CT/S$, and of the work function contribution $\Delta\phi^{CT}$ as a function of the interface Mg atom concentration.}
\begin{ruledtabular}
\begin{tabular}{cccc}
\multirow{2}{*}{Mg int.(\%)} & \multicolumn{3}{c}{3ML MgO/Ag-Mg}\\ \cline{2-4}
& $CT$ & $CT/S$ & $\Delta\phi^{CT}$\\
& ($e/cell$) & ($10^{-2}e/\AA^2$) & ($eV$)\\ \hline
0 & $+$0.21 & $+$0.59 & 0\\
25 & $+$0.02 & $+$0.04 & $+$0.25\\
50 & $-$0.15 & $-$0.43 &  $+$0.48\\
75 & $-$0.27 & $-$0.78  & $+$0.64\\
100 & $-$0.37 & $-$1.06  & $+$0.77\\
\end{tabular}
\end{ruledtabular}
\end{table}

One important result to be stressed at this stage is that the increase of the interfacial Mg concentration results in a charge transfer from the metal substrate to the oxide which tends to increase the metal/oxide work function. Therefore, the charge transfer mechanism is not sufficient by itself to explain the reduction of $\phi_m^*$ induced by the Mg incorporation in the Ag substrate. We can already argue that the work function changes related to the structural relaxation at the interface ($\Delta\phi^{SR}$), and/or to the electrostatic compression effect($\Delta\phi^{comp}$) must have to go against the charge transfer mechanism at the interface, i.e., must lead to a strong work function reduction of the MgO/Ag(001) system.

\subsection {MgO lattice distortion}

\begin{figure}[b]
\includegraphics[width=0.48\textwidth]{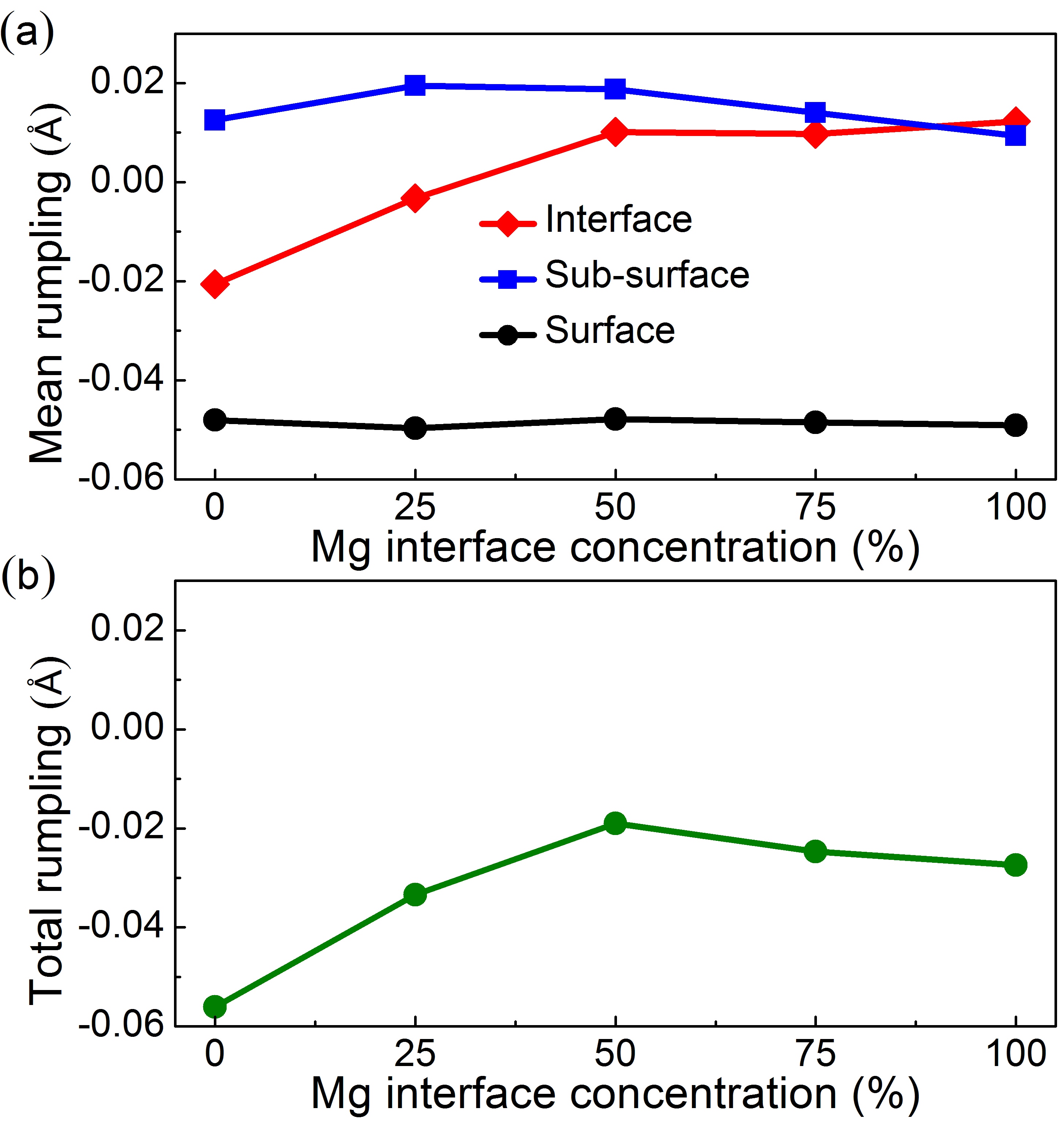}
\caption{\label{fig6}(color online) Evolutions of the mean rumpling of the interface, sub-surface and surface layers of the 3 ML thick MgO films (a) and of the oxide total rumpling (b) as a function of the interfacial Mg atom concentration.}
\end{figure}

As experimentally observed in section IV the incorporation of Mg atoms in substitutional sites of the Ag interface layer can induce a strong local distortion of the MgO interface layer. We expect that such a lattice deformation causes a modification of the electrostatic dipole in the MgO film which leads to work function changes of the system. We have thus studied the structural changes induced by the incorporation of interfacial Mg atoms in the MgO(3ML)/Ag(001) system. Fig. \ref{fig6}(a) shows the layer-by-layer evolution of the mean rumpling in each MgO layer deduced from the DFT-optimized geometry as a function of the Mg concentration. Here, the mean rumpling is defined as the average height difference between cations and anions for a given MgO layer. It is also shown on Fig. \ref{fig6}(b) the variations of the total rumpling in the oxide film.

We can see on Fig. \ref{fig6}(a) that the mean rumpling of the MgO interface layer strongly varies for Mg concentrations between 0\% and 50\% (+0.031 \AA) and then remains substantially unchanged. In the meantime, the variations for the sub-surface layer are much lower (standard deviation of 0.004 \AA~ around an average value of 0.015 \AA) and those of the surface layer are negligible (standard deviation lower than 0.001 \AA~ around an average value of -0.05 \AA). Thus, the dependence of the total rumpling on the interface Mg concentration is essentially related to the structural changes at the interface (84\% of the whole variation between 0\% and 50\% of interfacial Mg concentrations) (Fig. \ref{fig6}(b)).

For a non-zero rumpling in a given oxide layer, the negative and positive ions centers of charge do not coincide and this results in the presence of an electrostatic dipole. Consequently, a variation of the total rumpling within the MgO film leads to a change in the value of the electrostatic dipole through the insulator film, thereby modifying the work function. Therefore, since the total mean rumpling of the oxide film is changing with the Mg interfacial concentration, we expect a concomitant work function evolution. In order to obtain the rumpling contribution $\Delta\phi^{SR}$, we have studied the modifications of the MgO(3ML)/Ag(001) work function by varying the amplitude of the rumpling in the MgO surface layer around the equilibrium configuration. For a surface rumpling ($SR$) ranging from -0.05 \AA~ to +0.05 \AA~ we obtained a linear change of the work function as a function of the $SR$ with a slope of -9.90 $eV.\AA^{-1}$. The contribution $\Delta\phi^{SR}$ has been then obtained for all Mg atom concentrations by multiplying this slope value by the variation of the total mean rumpling of the MgO film. The results obtained for each Mg concentration are summarized in Table \ref{tab4}. 

\begin{table}[b]
\caption{\label{tab4}Dependence of the oxide film total rumpling $SR^{Tot}$ and of its work function contribution $\Delta\phi^{SR}$ to the whole MgO(3ML)/Ag(001) work function variation on the interface Mg atom concentration.}
\begin{ruledtabular}
\begin{tabular}{ccc}
\multirow{2}{*}{Mg int.(\%)} & \multicolumn{2}{c}{3ML MgO/Ag-Mg}\\ \cline{2-3}
& Mean rumpling ($SR^{Tot}$) & $\Delta\phi^{SR}$\\
& (\AA) & ($eV$)\\ \hline
0 & $-$0.056 & 0\\
25 & $-$0.033 & $-$0.23\\
50 & $-$0.019 & $-$0.37\\
75 & $-$0.025 & $-$0.31\\
100 & $-$0.027 & $-$0.29\\
\end{tabular}
\end{ruledtabular}
\end{table}

\begin{figure}[t]
\includegraphics[width=0.48\textwidth]{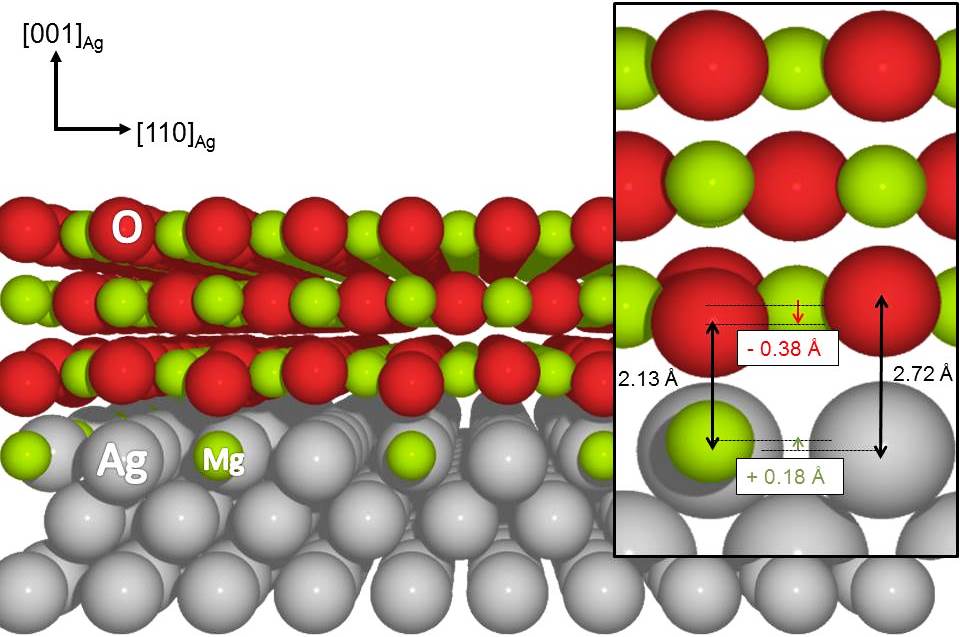}
\caption{\label{fig7}(color online). Sketch of the calculated atomic structure of the MgO(3ML)/Ag(001) system for an interface Mg atoms concentration of 25\% obtained after optimization. The inset shows the interface equilibrium distance $d_{Ag-O}=$2.72 \AA, the distance between the incorporated Mg atoms and the nearest O$^{2-}$ ions $d_{Mg-O}=$2.13 \AA, and the position variations of the Mg atoms contained in the Ag-Mg surface alloy (green color), and of the O atoms in the oxide interface layer (red color) with respect to their positions in the clean MgO(3 ML)/Ag(001) structure.}
\end{figure}

Let us now discuss the physical origin of the MgO local lattice distortion induced by the Mg incorporation. In Fig. \ref{fig7} we show a sketch of the DFT-optimized structure for 3ML MgO/Ag(001) with an interfacial Mg concentration of 25\%, a value close to the one experimentally obtained. We find that the essential of the local lattice distortion appears in the MgO interface layer. Although the equilibrium distance $d_{Ag-O}$ at the interface remains almost unchanged after Mg incorporation (2.72 \AA~ against 2.67 \AA~ before incorporation), the distance between the incorporated Mg atoms and the nearest O$^{2-}$ ions $d_{Mg-O}$ is highly reduced (2.13 \AA) with respect to $d_{Ag-O}$. Note that the bond length $d_{Mg-O}$ is close to the one experimentally found and to the Mg-O bond length in an MgO crystal. 

The very interesting point in this study is that the O$^{2-}$ ions directly above the Mg (Ag) atoms are displaced downwards (upwards) with respect to the other Mg$^{2+}$ ions by 0.29 \AA~ (0.09 \AA) in the MgO interface layer. As long as one neglects the effects related to the difference of polarisability of atoms and starting from a non-distorted configuration for the MgO/Ag-Mg system, this local structural deformation at the interface can be viewed as a direct consequence of Coulomb interactions between the positively charged Mg atoms (-1.5 electrons per atom as obtained from a Bader analysis) located in the Ag interface plane and the O$^{2-}$ ions directly adjacent. Besides, the Ag atoms neighboring the incorporated Mg atoms are negatively charged (0.9 electrons per atom as obtained from a Bader analysis) so that they exert a repulsive force on the nearest neighbors O$^{2-}$ ions. Together, the O$^{2-}$ ions exert an attractive (repulsive) force on the Mg (Ag) substrate atoms. 

As a result the Mg atoms in the Ag-Mg interfacial alloy are located slightly above the Ag atoms. For example the DFT-optimized geometry calculation gives an atomic corrugation amplitude of 0.18 \AA~ for a Mg concentration of 25\%. It can be noted that the in-plane charge transfer of the Mg atoms to the Ag atoms is also present even without the oxide over-layer (-1.2 electrons per atom) but in this case there is a very low corrugation in the Ag-Mg surface alloy. What can be learned here is that the local lattice distortion in the MgO interface layer is mainly a result of a charge transfer between the Mg and Ag atoms in the interfacial alloy \cite{Note2}.
 
\subsection {MgO-induced polarization effect}

The polarization effect that appears at any dielectric/metal interfaces is related to a reduction in the overspill electron density at the metal surface induced by the strong electrostatic field of the dielectric and is responsible for the strong reduction of the metallic substrate work function.\cite{Goniakowski2004, Prada2008, Giordano2006} In the case of the MgO/Ag(001) system, the so-called electrostatic “compression” effect,  leads to a reduction of the metal work function of -1.54 eV as calculated with VASP program (with the GPAW code we obtained -1.55 eV). \cite{Prada2008} In the present section, we determine the importance of the electrostatic “compression” effect in the work function reduction of the MgO/Ag-Mg/Ag(001) system as a function of the Mg concentration at the metal/oxide interface.

\begin{figure}[t]
\includegraphics[width=0.48\textwidth]{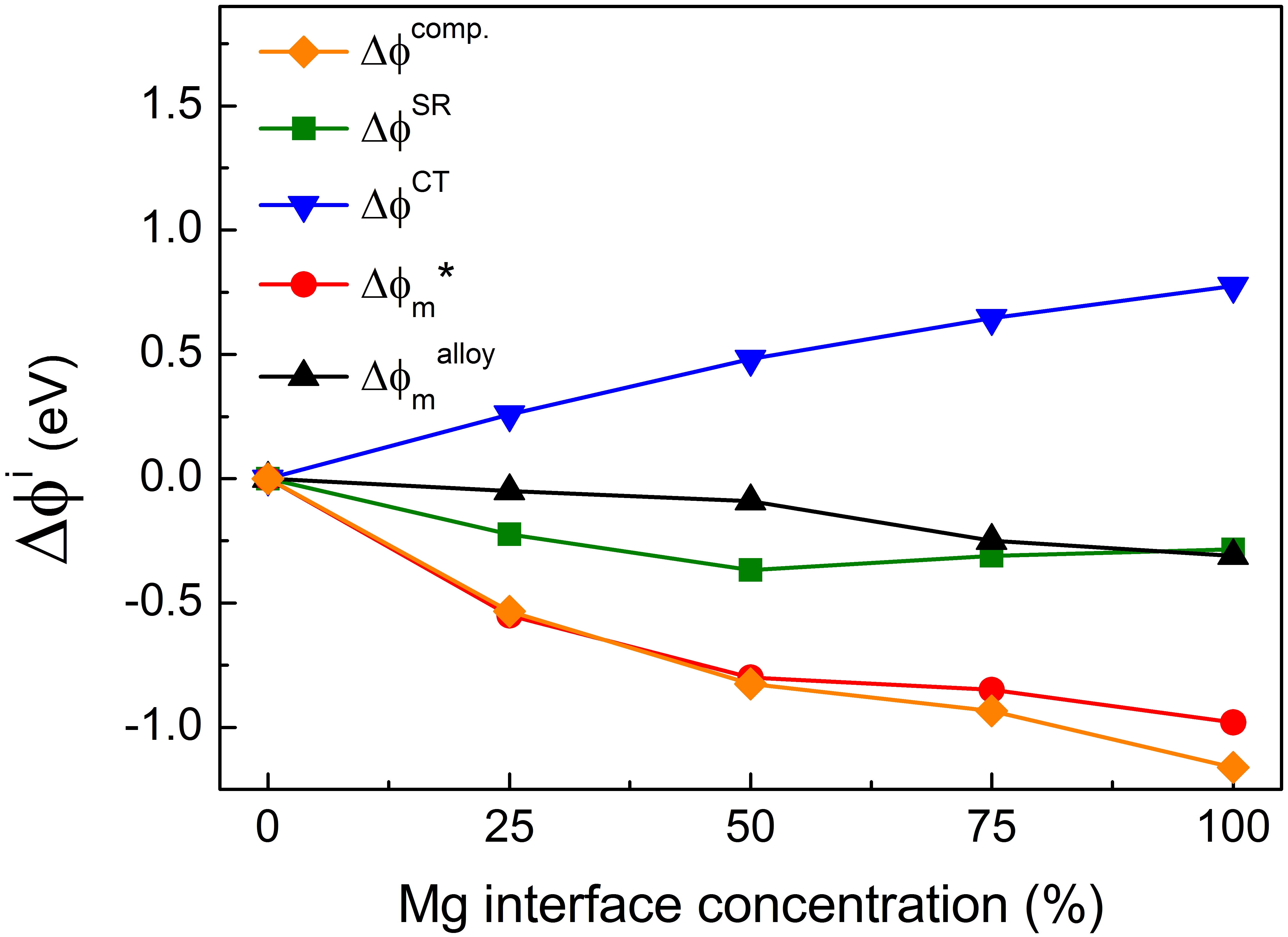}
\caption{\label{fig8}(color online). Evolutions of the calculated $\Delta\phi_m^{alloy}$, $\Delta\phi^{CT}$, $\Delta\phi^{SR}$, and $\Delta\phi^{comp}$ contributions to the calculated MgO(3ML)/Ag(001) work function variation $\Delta\phi_m^*$, as a function of the interface Mg atom concentration.}
\end{figure}

Up to now we have calculated the $\Delta\phi^{CT}$, $\Delta\phi^{SR}$, and $\Delta\phi_m^{alloy}$ contributions to the calculated MgO(3ML)/Ag(001) work function variation $\Delta\phi_m^*$ induced by the Mg atoms incorporation at the dielectric/metal interface. Knowing these different contributions an estimation of the “compression” effect as a function of the Mg concentration can be attempted using the following relation:
 
\begin{equation}
\Delta\phi^{comp}=\Delta\phi_m^*-(\Delta\phi^{CT}+\Delta\phi^{SR}+\Delta\phi_m^{alloy})
\end{equation}

Figure \ref{fig8} shows the evolutions of the calculated $\Delta\phi^{CT}$, $\Delta\phi^{SR}$, $\Delta\phi_m^{alloy}$, and $\Delta\phi^{comp}$ contributions to the calculated MgO(3ML)/Ag(001) work function variation $\Delta\phi_m^*$ as a function of the interface Mg atom concentration. The $\Delta\phi_m^*$ and $\Delta\phi^{comp}$ quantities decrease as a function of the Mg concentration and they are very close to each other in particular for the lowest Mg concentrations ($\leq$ 50\%), .i.e. the other contributions ($\Delta\phi^{CT}$, $\Delta\phi^{SR}$, and $\Delta\phi_m^{alloy}$) compensate each other. We can thus conclude that the decay of the metal/oxide work function is mainly governed by the progressive increase of the “compression” effect as a function of the Mg concentration that explains the work function reduction experimentally observed after the Mg incorporation at the oxide/metal interface. As we can see in Fig. \ref{fig8} the incorporation of 25\% of Mg in the first interface Ag layer results in theoretical work function change of -0.55 eV in very good agreement with our experimental findings.

In an \textit{ab initio} study of LiF/metal systems, it has been shown that the “compression” effect increases when the interface distance between the insulator and the metal decreases (the effect on the work function is typically of -1eV/\AA). \cite{Prada2008} For MgO(3ML)/Ag(001) we found that the interface distance only diminishes by 0.1 \AA~ when the Mg concentration of the interface Ag-Mg alloy varies from 0 to 100\% suggesting that the distance reduction effect induced by the Mg incorporation cannot itself explain the increase of the electrostatic “compression” contribution. In order to reach a deeper understanding of the observed effect with the MgO/Ag(001) system a more detailed analysis of the interface electronic structure is required as this was done for the BaO/metal systems by M. Nu{\~n}ez and M. B. Nardelli.\cite{Nunez2006}

\section {Conclusion}
   
In the present paper we have studied the mechanisms responsible for the work function changes induced by the Mg atoms incorporation at the MgO/Ag(001) interface by means of Auger electron diffraction experiments combined with multiple scattering calculation, \mbox{ultraviolet} photoemission spectroscopy, and density functional theory calculations. In our experimental conditions we demonstrate that Mg atoms are progressively and preferentially incorporated in the substitutional site of the Ag interface plane when the MgO films are exposed to a Mg flux. The incorporation of Mg atoms leads to a strong work function change (a reduction of 0.5 eV for an Mg concentration of 30\% in the Ag-Mg alloy) related to band-offset variations at the dielectric MgO/metal interface. We show that the Mg atoms in the interface Ag plane induce a strong local lattice distortion in the interface layers that mainly results in a charge transfer between the Mg and Ag atoms (electrons are transferred form Mg atoms toward Ag atoms) in the interfacial alloy combined with the Coulomb interaction between these atoms and their nearest neighbors O$^{2-}$ ions. The analysis of our DFT-calculation results indicates that this lattice distortion has only a limited impact on the work function reduction for the considered Mg concentrations. In fact the interesting aspect of our DFT-based study is that the work function reduction induced by the Mg incorporation at the MgO/Ag(001) interface is mainly governed by the electrostatic “compression” effect. This stems from the fact that the other contributions compensate each other. We believe that these findings give new insights for controlling the interfacial properties of dielectric/metal systems thereby providing the possibility of modulating the work function and the local structure of ultrathin oxide films surfaces which are known to have a significant impact on the reactivity properties of clusters/oxide systems in catalytic applications.\cite{Jung20112012, Jung2010} 
  
\begin{acknowledgments}
The authors warmly acknowledge A. Le Pottier and Y. Claveau for technical support. Parts of this work have been funded by European FP7 MSNano network under Grant agreement n$^{\circ}$ PIRSES-GA-2012-317554 and by COST Action MP1306 EUSpec. It has also been supported by the Fonds National Suisse pour la Recherche Scientifique through Division II.
\end{acknowledgments}

\end{document}